\documentclass[aps,prb,twocolumn,groupedaddress]{revtex4}
\usepackage{graphicx}

\begin{document}

\newcommand{\be}{\begin{equation}}
\newcommand{\ee}{\end{equation}}
\newcommand{\bea}{\begin{eqnarray}}
\newcommand{\eea}{\end{eqnarray}}
\newcommand{\nt}{\narrowtext}
\newcommand{\wt}{\widetext}

\title{Entanglement propagation through spin chains in the presence of a staggered magnetic field}

\author{R. H. Crooks and D. V. Khveshchenko}

\affiliation{Department of Physics and Astronomy, University of North
Carolina, Chapel Hill, NC 27599}

\begin{abstract}
We study the dynamics of entanglement in the $XY$ spin chain subject to a staggered magnetic field and contrast it to the previously studied uniform field case. We find that, depending on parameter values, a staggered field can provide better conditions for a perfect entanglement transfer, while even a modest amount of exchange anisotropy appears to have a strong detrimental effect. We also study interactions between different waves of entanglement and assess the possibility of simultaneous transmission of multiple bits of quantum information.
\end{abstract}

\pacs{03.67.Bg, 03.67.Hk} 
\keywords{keyword1, keyword2} 
\maketitle

A successful implementation of quantum computing protocols hinges on the possibility of achieving a high-fidelity transfer of quantum states between the different parts, such as 'core processor', 'storage', etc., of a quantum register.

To that end, a recently proposed variety of 'all-spin' architectures with 'always on' interactions has the potential of circumventing both the problem of conversion between physically different degrees of freedom and that of controling inter-qubit couplings individually.  

Specifically, the recent proposals have focused on the use of one dimensional chains of permanently coupled spin $1/2$-like variables ('stationary qubits') whose practical implementation might be possible with cold atoms in optical lattices, Josephson junction and quantum dot arrays, or other two-level systems. 

Being a salient feature of any periodic arrays, 
the translational invariance of spin chains
facilitates an emergence of quantum states with definite momentum ('flying qubits'), thereby enabling a high-fidelity spatial transfer of entanglement which is viewed as a vital resource for quantum computing.  

In most of the proposed designs, the local terms in the underlying Hamiltonian ('magnetic field') which can be used to perform one-qubit gates are generally expected to be much stronger than non-local 'exchange couplings' implementing two-qubit ones. Studied thus far are spin chains with ferromagnetic \cite{FM}, antiferromagnetic \cite{AFM}, short-range and long-range, as well as $SU(2)$- versus $U(1)$-invariant exchange couplings. On the other hand, much less attention has been paid to different patterns of the local field, almost all the studies being conducted in the (from the technical standpoint, simplest) case of a spatially uniform field.

Considering, however, that a local field can be used for both, initialization and readout of a quantum register, a comprehensive analysis of different local field distributions is strongly warranted. To that end, in this paper we study the time evolution of entanglement in the presence of a staggered field and contrast it to the previously explored uniform field case \cite{osterloh}. 

As an example of the system for which quantitative predictions can be made, we use the exactly soluble generalized anisotropic $XY$-model described by the $N$-spin Hamiltonian
\begin{equation}
H=\lambda\sum_{i=1}^N[(1+\gamma){S}^x_i{S}^x_{i+1}+(1-\gamma){S}^y_i{S}^y_{i+1}]-\sum_{i=1}b_i{S}^z_i.
\end{equation}
where $\gamma$ is the anisotropy parameter ranging between between $0$ (isotropic $XX$-model) and $1$ (Ising model). Furthermore, $\lambda=J/B$ is the nearest-neighbor exchange coupling and $b_i=B_i/B$ is the normalized local magnetic field in the $z$-direction, both in units of the average on-site field $B$. In what follows, we focus on the staggered, $b_i=(-1)^i$, field configuration and contrast it to the previously studied uniform, $b_i=1$, one.

The overall sign of the exchange term in Eq.(1) is chosen to be positive, although it can be readily inverted under a $\pi$-rotation ${\hat Z}=\prod_i{\hat R}^z_{2i+1}(\pi)$ about the $\hat z$-axis in the spin space on every other site:
$S^{x,y}_{2i+1}\rightarrow -S^{x,y}_{2i+1}$, $S^y_i\rightarrow S^y_i$, $S^{x,y}_{2i+1}\rightarrow -S^{x,y}_{2i+1}$, $S^{x,y}_{2i}\rightarrow S^{x,y}_{2i}$, $S^z_i\rightarrow S^z_i$.

Likewise, under a $\pi$-rotation ${\hat X}=\prod_i{\hat R}^x_{2i+1}(\pi)$ about the $\hat x$-axis the Hamiltonian of the $XY$-model in a staggered transverse field transforms into that in a uniform one, albeit at the expense of the signs' of the $\hat{x}$- and $\hat{y}$-spin couplings becoming opposite (note that in the Ising limit the two models become identical). Moreover, at arbitrary $\gamma\neq 0$ the rotated Hamiltonian can be brought back into the form of that in a uniform field by a simple rescaling: ${\hat X}{\hat H}(\lambda,\gamma,(-1)^i){\hat X}^{-1}={\hat H}(\lambda\gamma,1/\gamma,1)$.

Instead of resorting to the above (ill-defined in the isotropic limit $\gamma=0$) identity, we diagonalize the rotated Hamiltonian directly by applying the Jordan-Wigner fermion representation of the spin operators \cite{barouch}
\begin{equation}
\sigma^z_i=2c^\dagger_ic_i-1, \sigma^+_i=\prod_{j<i}(\sigma^z_j)c_i, \sigma^-_i=\prod_{j<i}(\sigma^z_j)c^\dagger_i.
\end{equation}
followed by the Bogoliubov transformation in the momentum space $\eta_k=u_kc_k - iv_kc^\dagger_{-k}$, with $u^2_k+v^2_k=1$, as a result of which the Hamiltonian takes the form 
$H=\sum_k\omega_k(\eta^\dagger_k\eta_k-\frac{1}{2})$.

In terms of the functions $A_k=1+\lambda\gamma\cos(k)$ and  $B_k= \lambda\sin(k)$, the Bogoliubov coefficients and fermion dispersion relation read $u_k =\sqrt{\frac{1-A_k/\omega_k}{2}}$, $v_k =\sqrt{\frac{1+A_k/\omega_k}{2}}$, and $\omega_k=\sqrt{A_k^2+B_k^2}.$

In the position space, the time evolution of the fermion operators is given by the equation of motion $\partial c_j(t)/\partial t=i\left[H,c_j(t)\right]$, whose solution is $c_j(t)=\sum_l\alpha_{l-j}(t)c_l -i\beta_{l-j}(t)c^\dagger_l$ with the time-dependent coefficients
\begin{equation}
\begin{array}{cc}
\alpha_j(t)=\frac{1}{N}\sum_k\cos(kj)[\cos(\omega_kt)-i
{A_k\over\omega_k}\sin(\omega_kt)],\cr
\beta_j(t)=\frac{1}{N}\sum_k\sin(kj){B_k\over \omega_k}\sin(\omega_kt)
\end{array}
\end{equation}
For $\gamma=0$ and $\lambda\ll1$ these expressions assume the form
\begin{equation}
\begin{array}{cc}
\alpha_j(t)\approx \left\{
\begin{array}{cc}
i^{-\frac{j}{2}}J_{\frac{j}{2}}(\frac{\lambda^2}{4}t) &\textit{j even}\cr
0 &\textit{j odd}
\end{array}\right.\cr
\beta_j(t)\approx\frac{\lambda}{N}\sum_k\sin(kj)\sin(k)\sin(\omega_kt)
\end{array}
\end{equation}

\begin{figure}[t]
	\centering
		\includegraphics{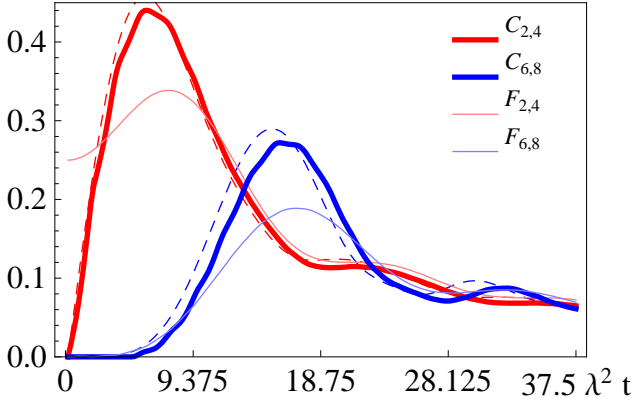}
	\caption{Concurrence $C_{i, i+2}$ and fidelity $F_{i, i+2}$ , $i$ sites removed from the initially entangled site at 
$(0,2)$ for the staggered field model with 
$\gamma=0$, $\lambda=0.05$. The dashed lines represent the results obtained for a uniform field.}
	\label{fig:figA}
\end{figure}

The above formulas allow one to determine the time evolution of an arbitrary initial state, although if the initial state is highly mixed this process becomes computationally rather cumbersome.  Therefore, in line with the previous research \cite{osterloh}
we focus upon simple initial spin configurations where all of the spin states are aligned with the field, with the exception that some pairs of spins begin in an entangled state.

In the staggered field model, we first look into the behavior of a state with only one singlet 
\begin{equation}
\psi_{i, j} = \frac{1}{\sqrt{2}}[c^\dagger_i + c_i -(c^\dagger_j + c_j)]\left|\uparrow\downarrow\uparrow\downarrow... \right\rangle.
\end{equation}
formed between the spins $i$ and $j$ residing on the 
same sublattice.

The entanglement properties of this state are described by a two-spin reduced density matrix $\rho^{(2)}_{m,n}$ which can be obtained from the density matrix of the entire system with all the sites but $i$ and $j$ traced out. In the basis $\left\{\left|\uparrow\uparrow\right\rangle, \left|\uparrow\downarrow\right\rangle, \left|\downarrow\uparrow\right\rangle, \left|\downarrow\downarrow\right\rangle\right\}$ it can be cast in the form:
\begin{equation}
\rho^{(2)}_{m,n}(t)=
\left(
\begin{array}{cccc}
\left\langle P^+_mP^+_n\right\rangle&\left\langle P^+_m\sigma^-_n\right\rangle&\left\langle  \sigma^-_mP^+_n\right\rangle&\left\langle \sigma^-_m\sigma^-_n\right\rangle \cr
\left\langle P^+_m\sigma^+_n\right\rangle&\left\langle P^+_mP^-_n\right\rangle&\left\langle \sigma^-_m\sigma^+_n\right\rangle&\left\langle \sigma^-_mP^-_n\right\rangle \cr
\left\langle \sigma^+_mP^+_n\right\rangle&\left\langle \sigma^+_m\sigma^-_n\right\rangle&\left\langle P^-_mP^+_n\right\rangle&\left\langle P^-_m\sigma^-_n\right\rangle \cr
\left\langle \sigma^+_m\sigma^+_n\right\rangle&\left\langle \sigma^+_mP^-_n\right\rangle&\left\langle P^-_m\sigma^+_n\right\rangle&\left\langle P^-_mP^-_n\right\rangle	
\end{array}
\right)
\end{equation}
where $P^\pm_m(t)=\frac{1}{2}(1\pm \sigma^z_m(t))$.  

Since $\psi_{i, j}$ and $H$ are both invariant under parity, the expectation values $\left\langle \sigma^x_m\right\rangle$, $\left\langle \sigma^y_m\right\rangle$, $\left\langle \sigma^x_m\sigma^y_n\right\rangle$, $\left\langle \sigma^x_m\sigma^z_n\right\rangle$, and 
$\left\langle \sigma^y_m\sigma^z_n\right\rangle$ vanish.

The remaining, non-vanishing, equal-time correlation functions appearing in Eq.(6) are most easily evaluated by rotating with the operator ${\hat X}$, so that the field becomes uniform and the initial state takes the form ${\hat X}\psi_{i, j} = \frac{1}{\sqrt{2}}(c^\dagger_i(0) - c^\dagger_j(0)) \left|\Downarrow \right\rangle$.

Thus it is necessary to evaluate matrix elements of the form $\left\langle \xi_{i, j}\right| {\hat X}F(P^\pm_m,P^\pm_n,\sigma^\pm_m,\sigma^\pm_n){\hat X}^{-1}\left|\xi_{i, j}\right\rangle$ which can be computed directly via Wick's theorem, or by a method involving sums of Pfaffians \cite{barouch,osterloh}.

As a convenient measure of pair-wise entanglement we will use concurrence defined according to the formula \cite{concurrence}
\begin{equation}
C_{i,j}=Max\left\{0, \lambda_4-\lambda_3-\lambda_2-\lambda_1\right\},
\end{equation}
where $\lambda_i$ are the eigenvalues $\rho^{(2)}_{i,j}\sigma_y\otimes\sigma_y(\rho^{(2)}_{i,j})^*\sigma_y\otimes\sigma_y$ in ascending order.  For a comparison, we will also compute fidelity defined as follows: $F^\psi_{i,j}=Tr(\rho^{(2)}_{i,j}\left| \psi \right\rangle \left\langle \psi\right|)$, where $\left|\psi\right\rangle$
is the singlet Bell state on the initially entangled pair of spins. 

\begin{figure}[t]
	\centering
		\includegraphics{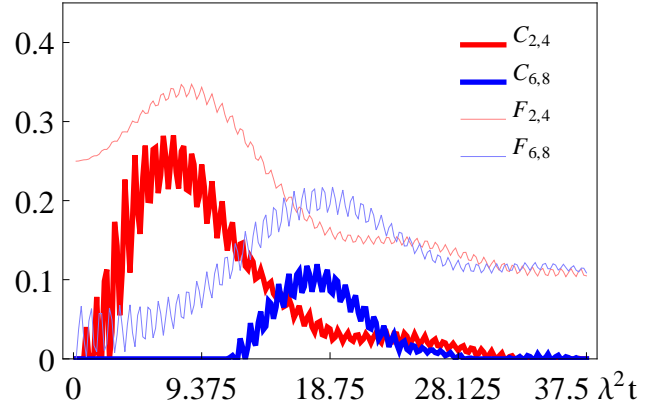}
	\caption{Concurrence $C_{i, i+2}$ and fidelity $F_{i, i+2}$ $i$ sites away from the initially entangled site at $(0,2)$ for the staggered field model with $\gamma=0$, $\lambda=0.5$.}
	\label{fig:figB}
\end{figure}

Starting out in the initial state (5), in the isotropic ($\gamma=0$) strong field ($\lambda\ll 1$) limit the system undergoes a time evolution described by Eq.(4) which can be viewed as the propagation of a singlet (EPR) pair which, after having been initially created at the sites $i$ and $j$, starts moving in both directions away from these sites at a speed ${\lambda^2}/{4}$.

Thus, consistent with the general analysis of Refs.\cite{magnon},
the entanglement propagation can be characterized 
in terms of the dispersion and damping (apparent decoherence rate) of underlying elementary excitations, as long as only a limited number of which is generated in the process of a quantum state transfer.

In the position space, the operator of a wave packet
corresponding to each of the two spreading 
entanglement waves is a superposition
\begin{equation}
 d_j \approx \sum_li^{(l-j)/2}J_{(l-j)/2}(\lambda^2t/4)d_l.
\end{equation}
which receives its largest contribution from the sites on the same sublattice as $i$, thereby manifesting the fact that the entanglement wave remains largely confined to one sublattice.

Figs. 1 and 2 show the results of the calculation of $C_{i,i+2}$ and $F_{i,i+2}$ for the isotropic ($XX$) staggered field model with $\lambda=0.05$ and $\lambda=0.5$, respectively.  In the plots, the initially entangled pair is on the sites numbered $0$ and $2$, labeled as $(0,2)$, while the fidelity and concurrence are computed on the sites $(2,4)$ and $(6,8)$.

In the strong field limit, the quality of entanglement transport in the staggered field model approaches that of the uniform field one
(the dashed lines).  
The apparent 'ripples' in the plots of Fig.1 should be attributed to vacuum fluctuations arising due to the fact that for any finite $\lambda$ the fully polarized (classical Neel) initial state differs from the true ground one.

With increasing $\lambda$ the vacuum fluctuations become more prominent and, concomitantly, the amount of transmitted entanglement decreases, as seen in Fig.2. While the singlet fidelity remains approximately constant, this reduction of the entanglement transfer is due to the emergence of additional states of the form 
$a\left|\uparrow\downarrow\right\rangle+b\left|\downarrow\uparrow\right\rangle$, whose contribution is governed by the coefficients $\beta_{l-i}\lambda$ from Eq.(4). 

Next, we study the time evolution of an initial state with two individual entangled EPR pairs
\begin{equation}
\xi_{i, j} = \frac{1}{2}(c^\dagger_i + c_i -(c^\dagger_j + c_j))(c^\dagger_k + c_k -(c^\dagger_l + c_l))\left|\uparrow\downarrow\uparrow\downarrow... \right\rangle
\end{equation}
Specifically, in Fig.3 we contrast the two cases when both pairs reside on the different (top picture, $(i,j),(k,l)=(24,26),(47,49)$ and same (bottom picture, $(i,j),(k,l)=(24,26),(46,48)$ sublattices, the brighter areas indicating greater values of the nearest sublattice-neighbor concurrence.  

The outward-moving entanglement waves propagate away from the initially entangled sites with the same speed $\lambda^2/4$ as that in the one-singlet case.  The lack of any detectable concurrence among nearest lattice-neighbor sites in the sublattice not containing the initially entangled sites is a direct consequence of Eq.(4), in the strong field limit.

The other two inward-moving wavefronts collide and then either pass through or annihilate each other, depending on which sublattice they reside.  If the initially entangled sites are on different sublattices, then according to Eq.(4) all of the fermion operators $c_j(t)$ that contribute to the time evolution of the system will have odd indices for one wave and even indices for the other. Therefore, in this case the time evolution of the two colliding waves proceeds independently.

If on the other hand, the two waves are on the same sublattice, Eqs.(4) and (8) may be used to show that, when the waves intersect, all entanglement they were transmitting is exactly negated by the other wave. This negative interference is a generic property of the uniform field models where it can present a serious problem for 
a simultaneous unimpeded propagation of multiple flying qubits. 

\begin{figure}[t]
	\centering
		\includegraphics{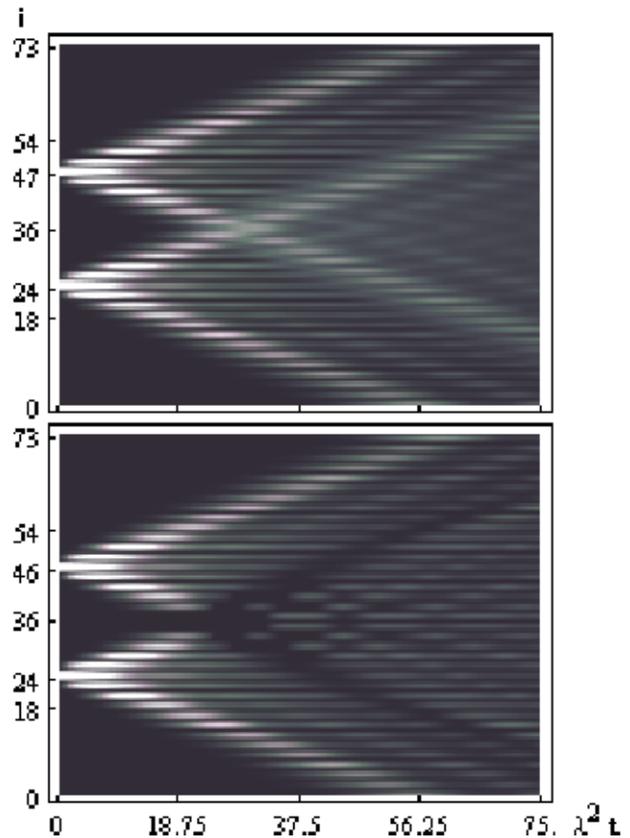}
	\caption{Time evolution of $C_{i, i+2}$ for $\lambda=0.05$ and $\gamma=0$.  The 2 initially entangled pairs of sites in the top plot are on different sublattices.  In the bottom plot the two pairs are on the same sublattice.}
	\label{fig:figC}
\end{figure}

As $\lambda$ is increased the term in Eq.(4) proportional to the function $\beta_j(t)$ not only serves to dampen the transmission as already mentioned, but also to diminish the degree to which the waves confined to different sublattices maintain their integrity. 
Formally, this occurs due to the function $\beta_j(t)$ taking a non-zero value for all values of $j$, which allows for the two waves to mix, resulting in their loss of coherence.

The observed nearly exact cancellation of the transmitted concurrence is not a general result, since it only occurs when the two colliding waves are created at precisely same time on the same sublattice. However, the ability of entanglement waves to pass through each other if they propagate on different sublattices is independent of the strength of the entanglement carried in each wave and also of the actual state that is transported.

Finally, we investigate the effect of a small anisotropy ($\gamma\ll 1$) in the strong (staggered) field limit ($\lambda\ll 1$)
where the fermion dispersion becomes $\omega_k\approx 1+\frac{\lambda^2}{4} + \gamma\lambda\cos(k)- \frac{\lambda^2}{4}\cos(k)$.  The mixed cosine terms do not allow for a closed form for the function $\alpha_j(t)$ unless $\lambda\gg\gamma$, in which case Eq.(4) still holds.

\begin{figure}[t]
	\centering
		\includegraphics{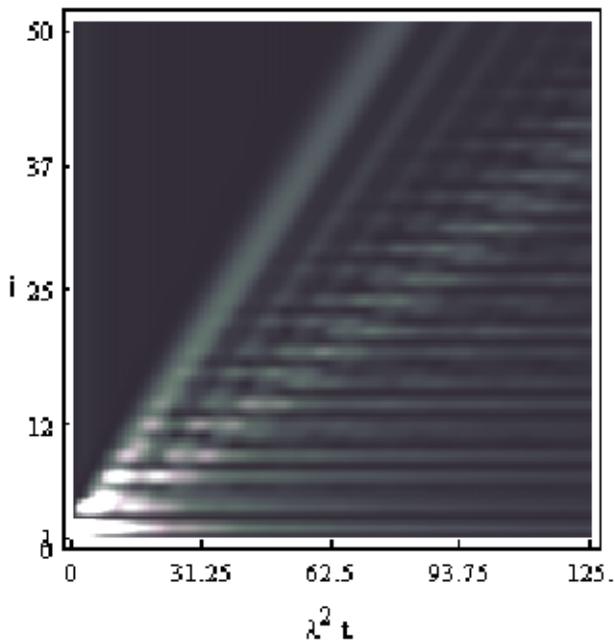}
		\caption{Time evolution of the concurrence $C_{i,i+2}$ for the anisotropic $XY$ system in a staggered field, $\gamma=0.0125$ and $\lambda=0.05$.}
	
	\label{fig:figD}
\end{figure}

In the opposite limit ($\lambda\ll\gamma$), one finds that $\beta_j(t)$ is again negligible, while the result
\begin{equation}
\alpha_j(t)\approx i^{-j}J_{j}(\gamma\lambda t).
\end{equation}
suggests that the initial excitation is no longer confined to its sublattice but propagates over the entire lattice, the corresponding wavefront traveling with the speed $\gamma\lambda$.

The intermediate regime $\lambda\sim\gamma\ll 1$ is presented in Fig.4
where the nearest sublattice-neighbor concurrence is plotted for only one side of the propagating wave in the case of $\gamma=0.0125$ and $\lambda=0.05$. The plot shows a presence of two types of elementary excitations, one propagating on all sites at a speed $\gamma\lambda$ and the other, with a speed $\lambda^2/4$, which is confined to one sublattice, as in the isotropic limit.

As the degree of anisotropy increases, there is a transfer of spectral weight from the latter to the former mode, which starts to dominate the transport of entanglement already at $\gamma\sim\lambda$ and continues all the way to the Ising limit.

In conclusion, we studied the dynamics of the spin $1/2$ $XY$ model in a staggered transverse field for a few simple out of equilibrium states. This previously unexplored problem is of interest in those prospective designs of a quantum register where the local field can be created by a $1D$ antiferromagnetic crystal (which systems happen to be more abundant than their ferromagnetic counterparts) placed in the vicinity of the $1D$ qubit array.

The use of non-uniform (especially, alternating) fields in qubit chains might also be advantageous in light of the their ability to create additional entanglement \cite{staggered}. Moreover, the study of various field configurations is important for optimizing local field profiles \cite{optimal} as a part of the general problem of finding the best parameter values for artificially engineered qubit chains (most notably, cold atoms in optical lattices or Josephson junctions arrays). 

To that end, we found that in the strong field limit the system approaches the uniform field model in its ability to transport entanglement, however, the speed of the state transfer appears to be lower because of an extra factor of the inverse field strength, which factor needs to be small in order to transport states with an acceptable degree of fidelity.

In this regime, the two sublattices of the chain act independently, so that waves traveling on one sublattice will not interfere with those on the other one. After their collisons, the integrity of such entanglement waves remains largely intact, thereby facilitating 
the possibility of transporting multiple quantum bits at the same time.

On the other hand, a generally slow speed of transmission in the staggered field case increases the propagating state's exposure to
any environmental sources of decoherence, thus manifesting yet another potential tradeoff which needs to be accounted for in the prospective designs of a practical $1D$ quantum register (of course, the emergence of mutually contradicting criteria is a rather common place in the field quantum computing in general and in the qubit-chain designs in particular \cite{dvk}).

Lastly, by examining the effect of anisotropy in the $XY$ staggered field model, we find that the transmission starts to lose the characteristics of staggered field propagation already for moderate values of the anisotropy parameter $\gamma$, as compared to the ratio of the coupling strength to the field strength $\lambda$.

Our results provide a further insight into the non-equilibrium 
dynamics of qubit chains and facilitate an ongoing quest for the optimal conditions of a reliable quantum information transfer.

This research was supported by NSF under Grant DMR-0349881.

\end{document}